**Spontaneous motion in hierarchically assembled active matter**

Tim Sanchez[1*], Daniel T. N. Chen[1*], Stephen J. DeCamp[1*], Michael Heymann[1,2] and Zvonimir Dogic[1]

[1]Martin Fisher School of Physics, Brandeis University, 415 South St., Waltham, MA 02454, USA

[2] Graduate Program in Biophysics and Structural Biology, Brandeis University, 415 South St., Waltham, MA 02454, USA

**With exquisite precision and reproducibility, cells orchestrate the cooperative action of thousands of nanometer-sized molecular motors to carry out mechanical tasks at much larger length scales, such as cell motility, division and replication[1]. Besides their biological importance, such inherently non-equilibrium processes are an inspiration for developing biomimetic active materials from microscopic components that consume energy to generate continuous motion[2-4]. Being actively driven, these materials are not constrained by the laws of equilibrium statistical mechanics and can thus exhibit highly sought-after properties such as autonomous motility, internally generated flows and self-organized beating[5-7]. Starting from extensile microtubule bundles, we hierarchically assemble active analogs of conventional polymer gels, liquid crystals and emulsions. At high enough concentration, microtubules form a percolating active network characterized by internally driven chaotic flows, hydrodynamic instabilities, enhanced transport and fluid mixing. When confined to emulsion droplets, 3D networks spontaneously adsorb onto the droplet surfaces to produce highly active 2D nematic liquid crystals whose streaming flows are controlled by internally generated fractures and self-healing, as well as unbinding and annihilation of oppositely charged disclination defects. The resulting active emulsions exhibit unexpected properties,**

**such as autonomous motility, which are not observed in their passive analogues. Taken together, these observations exemplify how assemblages of animate microscopic objects exhibit collective biomimetic properties that are starkly different from those found in materials assembled from inanimate building blocks, challenging us to develop a theoretical framework that would allow for a systematic engineering of their far-from-equilibrium material properties.**

Active materials are assembled from microtubule (MT) filaments, which are stabilized with the non-hydrolyzable nucleotide analog GMPCPP, leading to an average length of 1.5 μm. Bundles are formed by adding a non-adsorbing polymer, Poly(Ethylene Glycol) (PEG), which induces attractive interactions through the well-studied depletion mechanism. To drive the system far from equilibrium, we add biotin labeled fragments of Kinesin-1, a molecular motor which converts chemical energy from ATP hydrolysis into mechanical movement along a MT[8]. Kinesins are assembled into multi-motor clusters by tetrameric streptavidin, which can simultaneously bind and move along multiple MTs, inducing inter-filament sliding (Fig. 1a). In this respect, our experiments build upon important earlier work that demonstrated the formation of asters and vortices in networks of unbundled MTs and kinesin[9,10]. However, compared to these disperse networks, the proximity and alignment of depletion-bundled MTs greatly increases the probability of kinesin clusters simultaneously binding and walking along neighboring filaments, thus enhancing the overall activity.

Motor-induced sliding of aligned MTs depends on their relative polarity. Kinesin clusters generate sliding forces between MTs of opposite polarity, whereas no sliding force is induced between MTs of the same polarity[11-13]. To study dynamics of active bundles, a dilute suspension of MT bundles is confined to quasi-2D chambers. No sliding is observed for isolated bundles, indicating that they are locally polarity-sorted. However, due to passive diffusion, quasi-static bundles occasionally encounter each other and join due to attractive depletion interactions. Merging bundle domains are equally likely to have the same or opposite polarity, the latter case resulting in their relative sliding and even bundle disintegration (Fig. 1b). In one case, two polarity-sorted bundles meet with an initial anti-parallel orientation and promptly slide off each other. Subsequently, they rotate by 180° and rebind with relative orientation opposite to their original, and this time no sliding occurs (Supplementary Movie 1). Overall, active MT bundles almost exclusively exhibit extensile behavior in contrast to actin/myosin bundles which preferentially contract[14]. Consequently, as we show next, gels constituted from extensile MT bundles are strikingly different from actin/myosin gels, which exhibit bulk contraction[15,16].

Increasing the concentration of extensile MT bundles leads to the formation of percolating bundled active networks (BANs) with unique properties (Methods). For example, in contrast to classical polymer gels, which only respond passively to externally imposed stresses, BANs exhibit internally generated fluid flows (Fig. 1d, Supplementary Movie 2). These highly robust spatiotemporally chaotic flows occur throughout millimeter-sized samples and can persist for up to 24 hours, limited only by available chemical fuel (ATP) produced through the regeneration system. To gain insight into the microscopic dynamics that drive the formation of such patterns, we visualize constituent MT bundles at high magnification. Similar to the dilute case, we observe extension of bundles due to internal polarity sorting. However, in contrast to the dilute case, extending MT bundles are now connected to a viscoelastic network. This causes them to buckle at a critical length scale[17] and subsequently fracture into smaller fragments which quickly recombine with surrounding bundles of random relative polarity, generating further extension. Thus, BAN dynamics are determined by cycles of MT bundles undergoing polarity sorting, extension, buckling, fracturing and subsequent recombination (Fig. 1c, Supplementary Movie 3). The emergent patterns, involving thousands of MT bundles, are qualitatively similar to those observed in other non-equilibrium systems such as swarming bacteria or swimming spermatozoa[18,19]. Theoretical coarse-grained models of extensile rod suspensions predict the occurrence of hydrodynamic instabilities that may underpin such large-scale collective effects[20,21]. Being constructed from the bottom up, the properties of BANs are easily controlled and optimized. For example, kinesin walking velocity is tuned by ATP concentration, allowing us to examine how active flows depend on the underlying microscopic dynamics (Supplementary Movie 4)[8].

To quantify fluid flow, we embed micron-sized beads into the BANs (Fig. 2a, Supplementary Movie 5). The beads are coated with a polymer brush to suppress their depletion-induced binding to MT bundles. In the absence of ATP, the tracer particles probe a passive viscoelastic network, and their Mean Square Displacement (MSD) is sub-diffusive (Fig. 2b). At intermediate ATP concentrations, MSDs become super-diffusive at longer time scales, while remaining sub-diffusive at shorter timescales; indicating that the beads are being advected by internally generated flows. Similar effects are observed for beads suspended in active bacterial baths[22]. At saturating ATP concentrations, MSDs are essentially ballistic; i.e., beads move along straight paths. At longer timescales, all MSDs should become diffusive; however, this requires measuring long trajectories that are not accessible to our imaging system. We also examine the structure of internally generated fluid flow by calculating the spatial velocity correlation function between bead pairs, $\langle \vec{V}(R) \cdot \vec{V}(0) \rangle$, where R is their lateral separation (Fig. 2c). Correlation functions decay exponentially and become anti-correlated at approximately 200 μm. The spatial decay of $\langle \vec{V}(R) \cdot \vec{V}(0) \rangle$ taken at different ATP concentrations collapses onto a universal curve when rescaled by the peak velocity $\langle V(0)^2 \rangle$, revealing that the spatial structure of the internally generated fluid flow is independent of kinesin velocity (Fig. 2c). For saturating ATP concentrations, the average bead velocity (~2.2 μm/sec) is approximately three times the maximum kinesin velocity[8]. Previous work on unbundled MT/motor systems described temporally transient dynamics before settling into a quasi-static state consisting of either aster or vortex defects[10]. BANs are fundamentally different at both macroscopic and microscopic scales. At macroscopic scales, BANs exhibit steady state permanent flows that persist for days and are only limited by ATP availability and stability of the constituent proteins. At microscopic scales, BAN dynamics is driven by cascades of MT bundles extending, buckling, self-fracturing and self-healing. Such unique dynamics inherently demands the presence of a depletion agent, absent in previous studies. Third, BANs are very robust, exhibiting the same behavior over a wide range of microscopic control parameters (0.4% to 10% PEG, 0.5 to 5 mg/mL MTs, 10 μM to 3 mM ATP, 1.5 to 120 μg/ml kinesin). Finally, as we demonstrate next, BANs under confinement exhibit drastically different behavior from those of unbundled MT/kinesin networks[23].

It is well known that with increasing concentration, rod-like molecules undergo a transition to a nematic liquid-crystalline phase. To explore this regime we created a flat 2D oil-water interface stabilized with a surfactant, with BANs being dispersed in aqueous phase. Over time extensile MT bundles adsorb onto the PEG brush formed by the surfactant molecules, eventually covering the entire surface with a dense liquid crystalline monolayer of locally aligned bundles. The adsorbed 2D layer constitutes an active MT liquid crystalline phase, characterized by fast streaming flows and defect unbinding (Supplementary Movie 6). Further information about the nature of active MT liquid crystals can be garnered from examining the structure and dynamics of their defects. In general, active liquid crystals can have either nematic symmetry[24,25], as found in monolayers of amoeboid cells or vertically shaken monolayers of granular rods[26,27], or polar symmetry which is found in motility assays at high concentrations[5,28]. Polar liquid crystals form vortex and aster defects[5,29]. In contrast, we find that active MT liquid crystals form disclination defects of charge ½ or -½, implying the presence of nematic symmetry (Fig. 3a-c). This is expected since the basic building blocks of these materials are symmetrically extensile MT bundles. In equilibrium liquid crystals, defects are largely static structures whose presence is determined by either internal frustrations or external boundary conditions. In contrast, defects in active MT nematics exhibit unique spatiotemporal dynamics. They are created as uniformly aligned nematic domains extend, buckle, and internally self-fracture (Fig. 3d), similar to the dynamical cascades of MT bundles in 3D active networks. Once created, a fracture line terminates with a pair of oppositely charged disclination defects. Eventually, even as the fracture self-heals, the defects remain unbound and stream around until eventually annihilating with oppositely charged defects (Supplementary Movie 6). The rates of defect-creation and annihilation are balanced, creating steady-state streaming dynamics that persist for many hours. These observations exemplify how active nematics are fundamentally different from equilibrium ones, where fractures, internal streaming flows and spontaneous unbinding of defect pairs are never observed.

In a biological context, active fluids are frequently confined to the cytoplasm and it has been proposed that such confinement leads to emergence of coherent flows that can enhance cellular transport, a phenomenon known as cytoplasmic streaming[30,31]. For this reason we have encapsulated BANs in aqueous droplets emulsified in fluorinated oil. When squeezed between two surfaces, such water-in-oil active droplets exhibit an entirely unforeseen emergent property: persistent autonomous motility. The motility is highly robust, limited only by the sample lifetime, which is typically a few days (Fig. 4a, Supplementary Movie 7). Instead of moving along straight lines, the active emulsion droplets preferentially move in periodic patterns, with their average velocities reaching up to ~1 μm/sec. In comparison, droplets without any chemical fuel do not exhibit any motion (Fig. 4b). To elucidate the microscopic mechanism that drives droplet motility, we image the internal MT dynamics. For small droplets (less than 30 microns), MT bundles extend and push against the water-oil interface, reaching a quasi-static state. Unable to buckle, they remain dispersed throughout the droplet and no motility is observed. In contrast, when confined within larger droplets, active MT bundles exhibit dramatically different behavior. Similar to the flat surface, they adsorb onto the oil-water interface, exhibiting coherent spontaneous flows (Supplementary Movies 8, 9). When in frictional contact with a hard surface, these internal flows drive the motility of the entire droplet (Fig. 4c). Scanning through the bulk of these motile droplets reveals that their interior is largely devoid of MT bundles relative to their surface (Fig. 4d). The critical droplet size required for the formation of active liquid crystals is roughly similar to the collective buckling length scale that characterizes flows of 3D BANs. Spontaneous flows that drive droplet motility bear resemblance to cytoplasmic streaming observed in Drosophila Oocytes[30].

Equilibrium nematics confined to spherical surfaces exhibit interesting defect configurations that have been proposed as building blocks for assembly of higher-order equilibrium structures[32]. Our experiments provide a new opportunity to explore how dynamic defects of active nematics on spheres can be harnessed to control droplet motility. At the next level of hierarchy, they also enable future studies in which emulsions are used for studying synchronization, crystallization and jamming of self-propelled spheres at high concentrations[33].

Conventional polymer gels, liquid crystals and emulsions are quintessential systems of soft condensed matter physics. Studies of such materials have resulted in numerous fundamental advances and countless technological applications. However, the properties of these materials are limited by the laws of equilibrium statistical mechanics. Here starting with simple extensile bundles and culminating in the assembly of motile droplets, we have demonstrated the hierarchical assembly of active gels, liquid crystals and emulsions. Being liberated from equilibrium constraints, the tunable emergent properties of such far-from-equilibrium materials are strikingly different from their passive analogs, and as such, they can be used to guide the further development of the statistical mechanics of active matter.

**Methods Summary:** MTs were polymerized to a concentration of 6 mg/ml with GMPCPP. The non-hydrolyzing analogue of GTP reduces the MT nucleation barrier, resulting in very short filaments. Left at room temperature, the filaments anneal end-to-end, slowly increasing in average length. The behavior of active MT networks is very sensitive to length distribution of constituent filaments. For the experiments described here, after polymerization, we left MTs at room temperature for 2 days, at which point they had an average length of 1.5 μm. Biotin-labeled kinesin fragments (401 amino acids of the N-terminal motor domain of *D. melanogaster* kinesin) were purified from *E. Coli.* Motor clusters were assembled by mixing biotin-labeled K401 with tetrameric streptavidin at a molar ratio of 1.7:1. Active mixtures consisted of MTs, kinesin clusters, depleting polymer (PEG, 20kDa), anti-bleaching agents, and an ATP regenerating system. The regenerating system, used phosphoenol pyruvate (PEP) and pyruvate kinase/lactate dehydrogenase (PK/LDH) to maintain the ATP concentration at a constant level, allowing us to tune the kinesin velocity by controlling the initial ATP concentration. At highest PEP concentrations, the regenerating system kept samples active for tens of hours. Samples were observed in a conventional flow cell and surfaces were coated with a repulsive poly-acrylamide brush to suppress adsorption of proteins onto walls due to depletion or other non-specific interactions. The behavior of active samples was monitored by visualizing MTs with fluorescence microscopy, and by tracking 3 μm beads with brightfield microscopy. Beads were coated with poly-L-lysine-poly(ethylene glycol) (PLL-PEG) block copolymers, which suppressed their sticking to the MT bundles. Emulsions were made with a PFPE-PEG-PFPE surfactant (E2K0660) in 3M HFE7500 oil.

**Acknowledgements:** We acknowledge insightful discussions with Robert Meyer, Robijn Bruinsma and Erwin Frey about the nature of liquid crystalline defects. Biotin-labeled Kinesin 401 (K401) was a gift from Jeff Gelles. This work was supported by the W. M. Keck Foundation, the National Institute of Health (5K25GM85613), the National Science Foundation (NSF-MRSEC-0820492, NSF-MRI 0923057) and the Pioneer Research Center Program through the National Research Foundation of Korea (2012-0001255). We acknowledge use of the MRSEC Optical Microscopy facility.

**Author Information** Reprints and permissions information is available at www.nature.com/reprints. The authors declare no competing financial interests. Readers are welcome to comment on the online version of this article at www.nature.com/nature. Correspondence and requests for materials should be addressed to Zvonimir Dogic (email: zdogic@brandeis.edu)

*These authors contributed equally to this work.

**Author Contributions** T. S., D. C., S. D. and Z. D. conceived of the experiments and interpreted the results. T. S., D. C. and S. D. performed the experiments. T. S. and D. C. conducted data analysis. T. S., D. C. and Z.D. wrote the manuscript. M.H. synthesized surfactant and contributed to assembly of active emulsions.

**Fig. 1 Active MT networks exhibit internally-generated flows.** (**a**) Schematic illustration of an extensile MT-kinesin bundle, the basic building block used for the assembly of active matter. Kinesin clusters exert inter-filament sliding forces while depleting PEG polymers induce MT bundling. (**b**) Two MT bundles merge and the resultant bundle immediately extends, eventually falling apart. Time interval, 5 seconds, 15 μm bar. (**c**) In a percolating MT network, bundles constantly merge (red arrows), extend, buckle (green dashed lines), fracture, and self-heal to produce a robust and highly dynamic steady-state. Time interval, 11.5 seconds, 15 μm bar. (**d**) An active MT network viewed on a large scale. Arrows indicate local bundle velocity direction. 80 μm bar.

**Fig. 2. ATP concentration controls dynamics of active MT networks.** (**a**) Tracer particles embedded in BANs indicate local fluid flow. Trajectories of the particles (red paths) reveal highly non-Brownian motion. 80 μm bar. (**b**) Mean square displacements of tracer beads plotted as a function of time for different ATP concentrations. The exact ATP concentrations, increasing from blue to red, are indicated in (c). (**c**) Normalized spatial velocity-velocity correlation functions as a function of lateral separation for varying concentrations of ATP. The velocities were determined using a time interval of 5 seconds. When normalized by the peak velocity $\langle V(0)^2 \rangle$, the correlation functions rescale onto a universal curve, revealing a characteristic lengthscale that is independent of ATP concentration. Inset: Bare spatial correlation functions reveal that average velocity depends on ATP concentration.

**Fig. 3. Dynamics of 2D streaming nematics confined to fluid interfaces. (a)** Schematic illustrations of the nematic director configuration around disclination defects of charge ½ and -½. **(b-c)** Active liquid crystals exhibit disclinations of both ½ and -½ charge, indicating the presence of nematic order. 15µm bar. **(d)** A sequence of images demonstrates buckling, folding and internal fracture of a nematic domain. The fracture line terminates with a pair of oppositely charged disclination defects. After the fracture line self-heals, the disclination pair remains unbound. 15 second time lapse. 20µm bar.

**Fig. 4. Motile water-in-oil emulsion droplets. (a)** Droplets containing extensile MT bundles exhibit spontaneous autonomous motility, when partially compressed between chamber surfaces. A droplet trajectory taken over a time interval of 33 minutes is overlaid onto a brightfield droplet image. **(b)** In the absence of ATP, passive droplets exert no internal forces, and the only contribution to their movement is minor drift. 80 µm bar. **(c)** Fluorescence image of active MT bundles which spontaneously adsorb onto the oil-water interface. The resulting active liquid crystalline phase exhibits streaming flows, indicated with blue arrows. Red arrow indicates instantaneous droplet velocity. The image is focused on the droplet surface that is in contact with the coverslip. 100 µm bar. **(d)** Image of the droplet taken at a midplane indicates that the droplet interior is largely devoid of MT bundles. 100 µm bar.

Figure 1

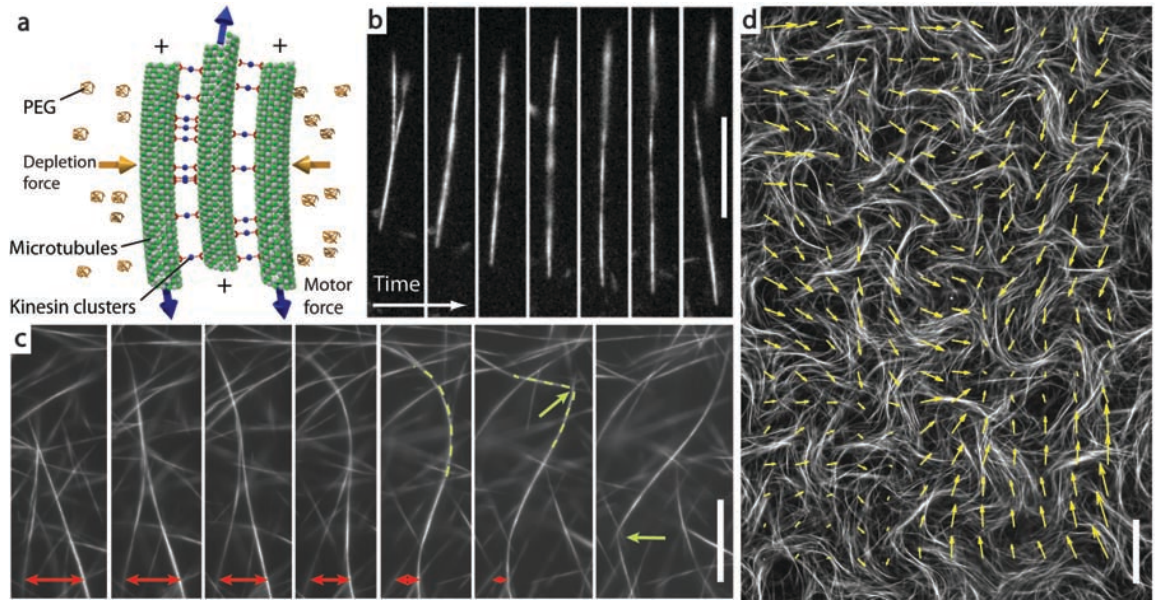

Figure 2

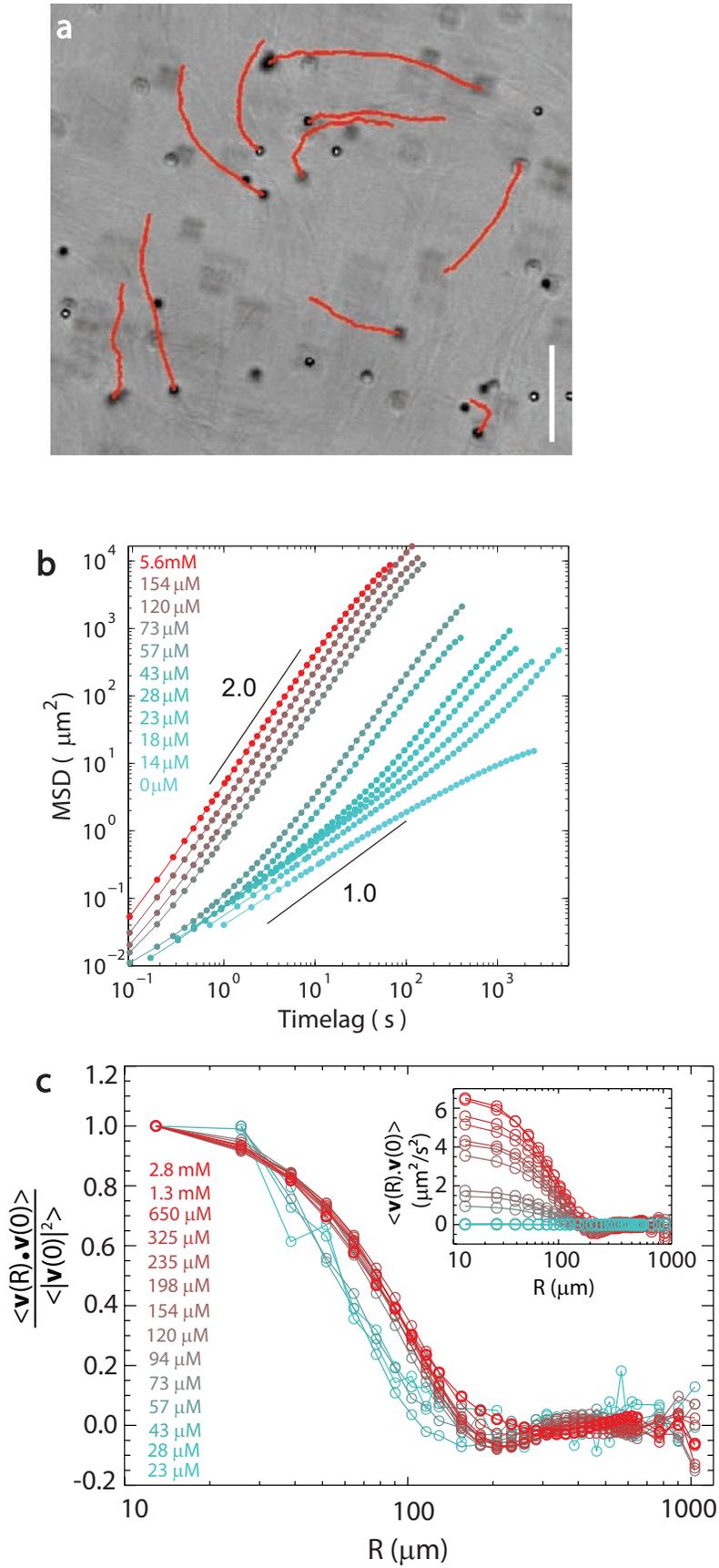

Figure 3

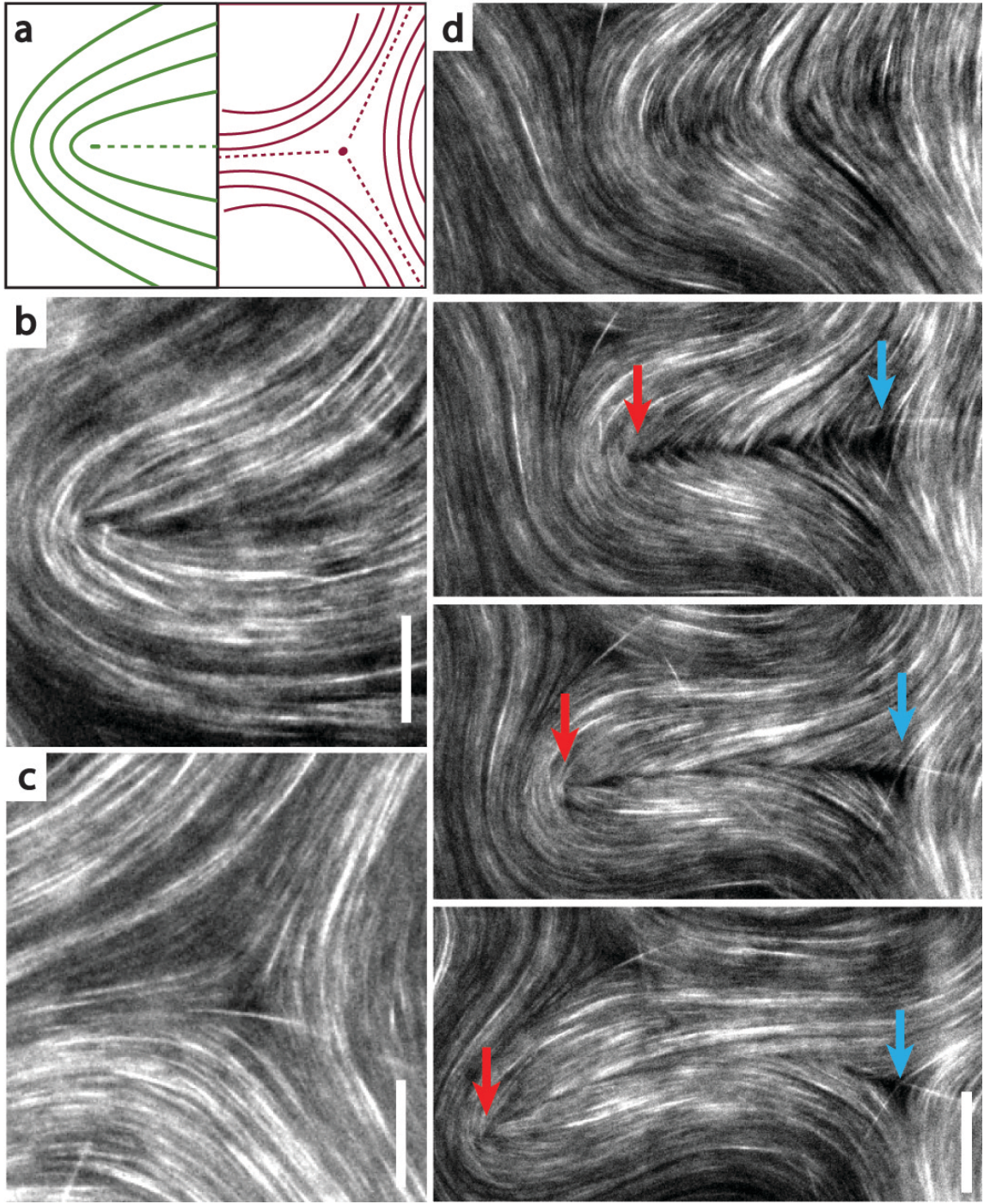

Figure 4

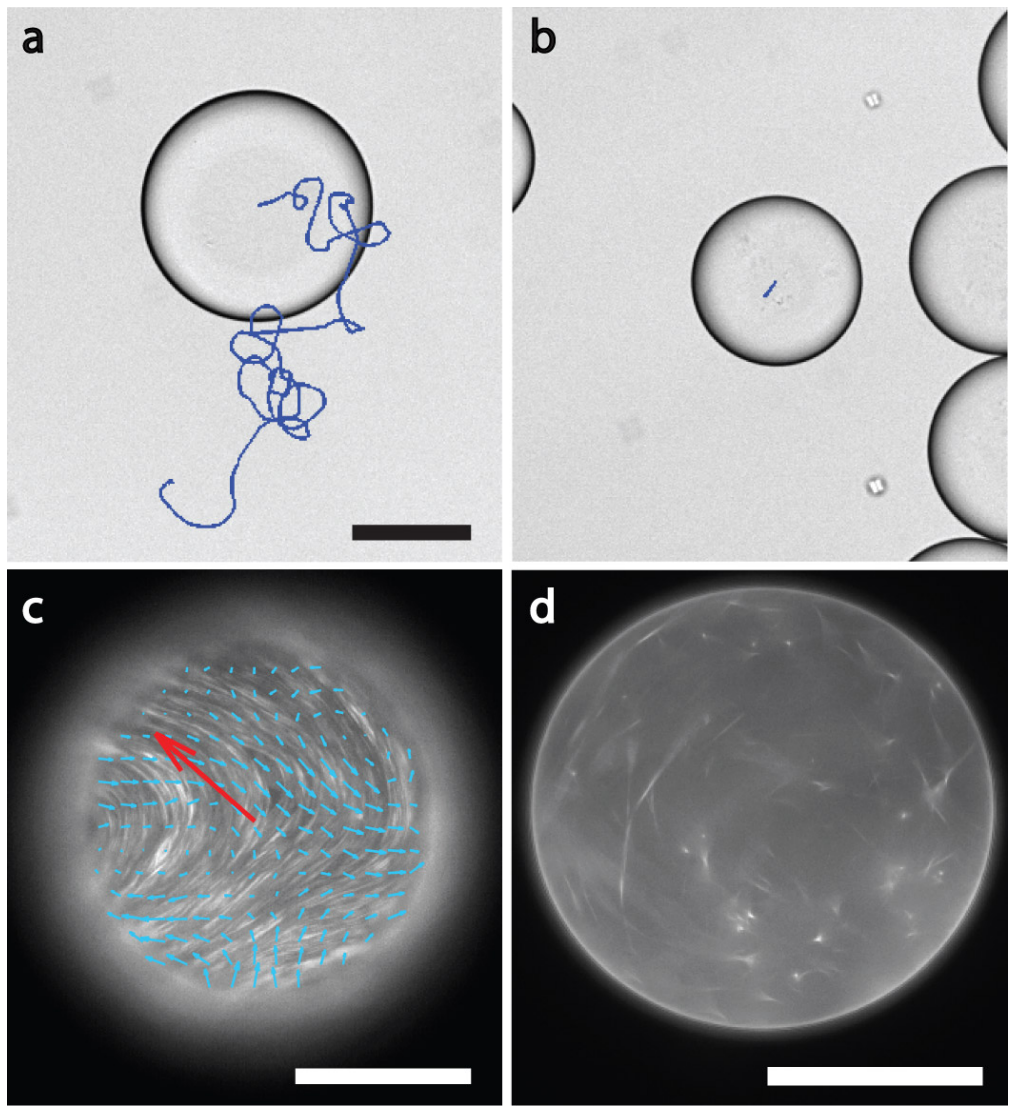

**Supplemental Videos available at:**

http://www.nature.com/nature/journal/v491/n7424/abs/nature11591.html - /supplementary-information

Methods

**Tubulin purification and MT polymerization:** Tubulin was purified from bovine brain through two cycles of polymerization-depolymerization in high-molarity PIPES buffer[34]. The purified protein had a concentration of 7.4 mg/ml in M2B buffer (80 mM PIPES pH 6.8, 1 mM EGTA, 2 mM $MgCl_2$) and was stored at -80 °C. Tubulin was subsequently recycled and flash-frozen at a concentration of 13.2 mg/ml in liquid nitrogen using thin-walled PCR tubes. For fluorescence microscopy, tubulin was labeled with Alexa Fluor 647 (Invitrogen, A-20006) by a succinimidyl ester linker; absorbance spectroscopy indicated that 29% of monomers were labeled[35]. Recycled tubulin was copolymerized with fluorescently-labeled tubulin producing MTs with 3% of labeled monomers. The polymerization mixture consisted of 4.5 μL (13.2 mg/ml) unlabeled tubulin, 0.9 μL (7.4 mg/ml) Alexa-647-labeled tubulin, 1.4 μL glycerol (to 15% of final), 0.56 μL GMPCPP (10mM)[35] (Jena Biosciences, NU-4056), 0.47 μL DTT (20 mM), and M2B buffer to a final volume of 9.4 μL (for a final tubulin concentration of 6.8 mg/ml). The suspension was incubated for 30 minutes at 37°C, and subsequently diluted to 6 mg/ml with M2B. Microtubules were annealed at room temperature for 2 days before using them for mixing experiments. This resulted in an average microtubule length of 1.5 μm.

**Kinesin-streptavidin complexes:** K401, which consists of 401 amino acids of the N-terminal motor domain of *D. melanogaster* kinesin, was purified as previously published[36]. The protein was frozen at 0.7 mg/ml in 50 mM imidazole (pH 6.7), 4 mM $MgCl_2$, 2 mM DTT, 50 μM ATP and 36% sucrose buffer. Kinesin-streptavidin complexes were assembled by mixing 7 μL of freshly thawed K401 solution containing 3 mM DTT with 8 μL of 0.35 mg/ml streptavidin (Invitrogen, S-888). The mixture was incubated on ice for at least 10 minutes before diluting with M2B to a final volume of 28 μL.

**Preparation of polymer-coated sterically repulsive microspheres:** Coating microspheres with a polymer brush leads to a steric repulsion that prevents their adsorption onto surfaces[37]. Negatively charged polystyrene beads were coated with a block-copolymer (PLL-PEG) consisting of poly-L-lysine backbone and poly(ethylene glycol) side chains. The positively-charged lysine groups electrostatically attach to the bead surface, leaving the PEG chains extending off the surface. PLL-PEG is synthesized by conjugating *N*-hydroxysuccinimidyl ester of methoxypoly(ethylene glycol) propionic acid (Laysan Bio, MPEG-SCM-20K-1g, MW 20kDa) with poly-g-lysine (Sigma, P7890-100MG, MW 20kDa) in 50mM sodium borate buffer, pH 8.5. The concentration of lysine amino acids is 10mM and the concentration of 20kDa PEG molecules is 2.86 mM. This mixture is incubated at room temperature with gentle mixing for an hour and is subsequently transferred into a dialysis membrane (pore size 5kDa). It is dialyzed two times against dH20, each time for more than 8 hours.

PLL-PEG block-copolymer was added to 3 μm polystyrene beads decorated with sulfate groups in a low salt buffer. The final mixture containing 10 μL PLL-PEG, 100 μL HEPES (100mM, pH 7.5), 12 μL bead stock (8% w/V) and 878 μL $H_2O$ was incubated for 1 hour, stirred gently and sonicated every 20 minutes. Subsequently, the beads were centrifuged for 5 minutes (5000 g) and resuspended in M2B buffer at ~1.3% w/v. Beads were imaged using brightfield microscopy and their positions were tracked using well-established protocols[38].

**Flow cell construction and sterically repulsive chamber surfaces:** Microscope glass slides and coverslips were coated with a poly-acrylamide brush to prevent non-specific adsorption of protein[39]. Slides were first rinsed and sonicated with hot water containing 0.5% detergent, then with ethanol, and finally with 0.1 M KOH. Subsequently, the slides were soaked in a mixture of 98.5% ethanol, 1% acetic acid, and 0.5% of the silane-bonding agent 3-(Trimethoxysilyl) propylmethacrylate (Acros Organics, 216551000) for 15 minutes. Slides were rinsed a final time and immersed in a 2% w/v aqueous solution of acrylamide. 35 µL/100 ml of TEMED and 70 mg/100 ml of ammonium persulfate were added to the acrylamide solution to promote polymerization of poly-acrylamide brush that is covalently attached to the glass surfaces. Slides and coverslips were stored in a suspension of polymerized acrylamide and used for up to 2 weeks. Each slide was rinsed and air-dried immediately before use.

**Assembly of active MT bundles:** To study the dynamics of dilute MT bundles we confined a low concentration of MTs to a quasi-2D observation chamber. In the absence of molecular motors, depletion interaction causes formation of MT bundles with mixed polarity. Introducing kinesin clusters induces relative sliding and eventual disintegration of bundles in which MTs have mixed polarity while leaving intact bundles in which MTs have the same polarity. This leads to a steady state which consists only of bundles with same polarity. In the dilute limit (0.25 mg/ml) the average bundle length is 9 µm while the average cross-section contains approximately 10-20 MT bundles. Similar bundles thicknesses are measured in 3D bundled active networks.

**Preparation of active microtubule networks:** Several initial mixtures were prepared separately and then combined into the final solution. Individual components were dissolved and frozen separately in M2B buffer, and then thawed freshly for use. We included anti-oxidant components

(listed below) and trolox (Sigma, 238813) to avoid photo-bleaching during fluorescence imaging. Anti-oxidant solution 1 (AO1) contained 15 mg/ml glucose and 2.5 M DTT. Anti-oxidant solution 2 (AO2) contained 10 mg/ml glucose oxidase (Sigma G2133) and 1.75 mg/ml catalase (Sigma, C40). Second, a high salt M2B solution (69 mM $MgCl_2$) was included to raise the $MgCl_2$. Lastly, ATP-regenerating components, including enzyme mixture pyruvate kinase/lactate dehydrogenase (PK/LDH, Sigma, P-0294) and phosphoenol pyruvate (PEP) were included in the final mixture. As ATP is hydrolyzed by kinesin activity, the PK/LDH uses PEP as a fuel source to convert ADP back into ATP at a rate that is much faster than the ATP hydrolysis. Thus constant ATP concentration is maintained throughout the experiments until all the PEP is exhausted[40].

The components described above were mixed without microtubules and in large volumes to reduce pipetting inaccuracies, while not wasting valuable tubulin. The "active pre-mixture" contained the following: 1.3 μL AO1, 1.3 μL AO2, 1.7 μL ATP (variable initial mM), 1.7 μL PK/LDH, 2.9 μL high-$MgCl_2$-M2B, 3 μL of sterically repulsive beads (3 μm beads, ~1.3% w/V), 6 μL trolox (20 mM), 8 μL PEP (200 mM), 8 μL PEG (6% w/w in M2B), 4 μL kinesin-streptavidin mix, and 12.1 μL M2B buffer. To prepare the final active mixture, 5 μL of the active pre-mixture was mixed with 1 μL of the 6 mg/ml microtubule solution, resulting in a final microtubule concentration of 1 mg/ml.

**Assembly of 2D active MT nematics at oil-water interface:** A 90 μm poly-acrylamide flow cell was constructed as described above. HFE7500 oil with 2% (V/V) PFPE-PEG-PFPE surfactant was flowed into the cell and immediately displaced by flowing in the active microtubule mixture. The oil preferentially wets the interface leaving a perfectly stable flat 2D oil-water interface decorated with the surfactant. The BANs were allowed to adsorb onto the oil-water interface and then imaged with fluorescent microscopy. Experiments indicate that the PEG is necessary for the adsorption of the MT bundles onto the surfactant monolayer and that active nematics assemble on flat as well as curved surfaces. Furthermore, the presence of surfactant is essential for the preservation of the streaming state and adsorption is only observed in active samples. In the absence of ATP, bundles do not adsorb onto the interface, instead forming a bulk rigid gel.

**Light microscopy:** Active MT networks were viewed with epi-fluorescence microscopy (Nikon Eclipse Ti microscope). Alexa Fluor 647-labeled microtubules were illuminated with a 120W metal halide light source (X-cite 120) and a fluorescent filter cube (Semrock, Cy5-4040B-NTE). Images were acquired with a cooled CCD camera (Andor Clara). To view a large sample area, we used the motorized stage and the Ti's Perfect Focus system to acquire adjacent fields of view, which were subsequently stitched together using a MatLab routine. 3µm tracer particles were imaged with standard brightfield microscopy, and particle tracking was performed with Matlab software[38]. Particle image velocimetry of the bundle flows was performed using the freely available Matlab based package PIVlab (version 1.11).

**Active mixtures in aqueous emulsion droplets:** The active MT network in aqueous suspension was mixed with 3M HFE7500 oil ENREF_21 at a 1:5 ratio and vortexed briefly 5 times to create water droplets within the oil medium. The aqueous droplets were stabilized with a 2% v/v solution of PFPE-PEG-PFPE surfactant (E2K0660) ENREF_35 [41]. Glass slides and cover slips were cleaned using the protocol previously described for preparation of poly-acrylamide coated slides. The emulsion was co-flowed into the microscope chamber with extra HFE7500 oil to disperse the droplets. Within the droplets, the dynamics of MT bundles were viewed with fluorescence microscopy, while the motion of the droplets themselves was examined with brightfield microscopy.